\documentclass[prl,twocolumn,superscriptaddress,showpacs]{revtex4}
\usepackage{graphicx}
\usepackage{epsfig}
\usepackage{amssymb,amsmath,amsfonts,hyperref}
\usepackage{wasysym}
\usepackage{latexsym}
\usepackage{eucal}
\usepackage{verbatim}
\usepackage[normalem]{ulem}

\usepackage{color}

\newcommand{\be}{\begin{eqnarray}}
\newcommand{\ee}{\end{eqnarray}}

\begin{document}
\title{Sign-problem-free Monte Carlo simulation of certain frustrated quantum magnets}
\author{Fabien Alet}
\affiliation{{\small Laboratoire de Physique Th\'eorique, IRSAMC, Universit\'e de Toulouse, CNRS, 31062 Toulouse, France}}
\author{Kedar Damle}
\affiliation{{\small Department of Theoretical Physics, Tata Institute of Fundamental Research, Mumbai 400 005, India}}
\author{Sumiran Pujari}
\affiliation{{\small Department of Physics \& Astronomy, University of Kentucky, Lexington, KY-40506-0055}}
\date{\today}

\begin{abstract}
We introduce a Quantum Monte Carlo (QMC) method for efficient sign-problem-free simulations of a broad class of frustrated $S=1/2$ antiferromagnets using the basis of spin eigenstates of clusters to avoid the severe sign problem faced by other QMC methods. We demonstrate
the utility of the method in several cases with competing exchange interactions,
and flag important limitations as well as possible extensions of the method.
\end{abstract}

\vskip2pc

\maketitle
{\em Introduction.-} Quantum Monte Carlo (QMC) simulations compute equilibrium properties of a many-body system by importance sampling of the canonical partition function
$Z ={\rm Tr} \exp(-H/T)$, where $H$ is the many-body Hamiltonian, and $T$ is the temperature~\cite{Stronglycorrelated_book,Krauth_book,Sandvik_review}.
They have emerged as a major tool for the study
of lattice Hamiltonians that either model low-$T$ thermodynamic properties of interesting strongly correlated materials~\cite{Johnston,frustration_book}, or provide concrete realizations
of novel phases in such condensed matter systems~\cite{Wessel_review,Kaul_review}. 
However, models of {\em geometrically frustrated magnets}~\cite{frustration_book}, in which antiferromagnetic interactions compete with each other due to the geometry of the exchange pathways, have
typically remained beyond the reach of QMC methods. This is due to the presence of a {\em sign problem}, whereby the weights
assigned to individual Monte Carlo configurations are no longer strictly positive in the commonly used basis of eigenstates of $S^z_{\vec{r}}$, the $z$-component of each spin.
In such cases, the average sign decreases exponentially with system size and inverse temperature, leading to unmanageably
large statistical errors in the estimation of physical quantities. A similar sign-problem
crops up in diverse settings ranging from QCD to strongly-correlated metals, and a general
solution is considered unlikely~\cite{Troyer_Wiese}.

Limited progress has been possible in a few cases, for instance in anisotropic systems in which the frustration only affects $S^z_{\vec{r}}$ (thereby allowing sign-free
simulation in the $z$ basis~\cite{Moessner_Sondhi,Isakov_Moessner,Wessel_Troyer,Heidarian_Damle,Melko_etal,Isakov_etal,Isakov_Kim_Paramekanti,Banerjee_etal,Sen_Damle_Senthil}), or when the sign problem of the full theory can be finessed at low T by working with a low-energy effective Hamiltonian 
which has no sign problem~\cite{Damle_Senthil,Sen_etal,Sen_Damle_Vishwanath,Berg,Huang_Chen_Hermele}. For some models, sign-free simulations are possible by virtue of specific symmetries of the Hamiltonian~\cite{Assaad_Evertz,Wu_Zhang,Messio,Huffman,Wang,Li_etal,Wei_etal}. In other strongly correlated systems with a full-fledged sign problem, progress has been made in some cases by developing {\em improved estimators} for computing physical observables~\cite{Sandvik_Henelius,Bietenholz_Pochinsky_Wiese,Chandrasekharan_Wiese,Nyfeler}. In principle, one could also try to find an alternate basis in which all configurations have positive weights. However, this has been possible only in a few interesting cases~\cite{Nakamura,Okunishi},  including some models of topologically ordered states of matter~\cite{Geraedts_Motrunich_ED, Geraedts_Motrunich_TI,Geraedts_Motrunich_FQH}.
 
\begin{figure}
{\includegraphics[width=0.95\columnwidth]{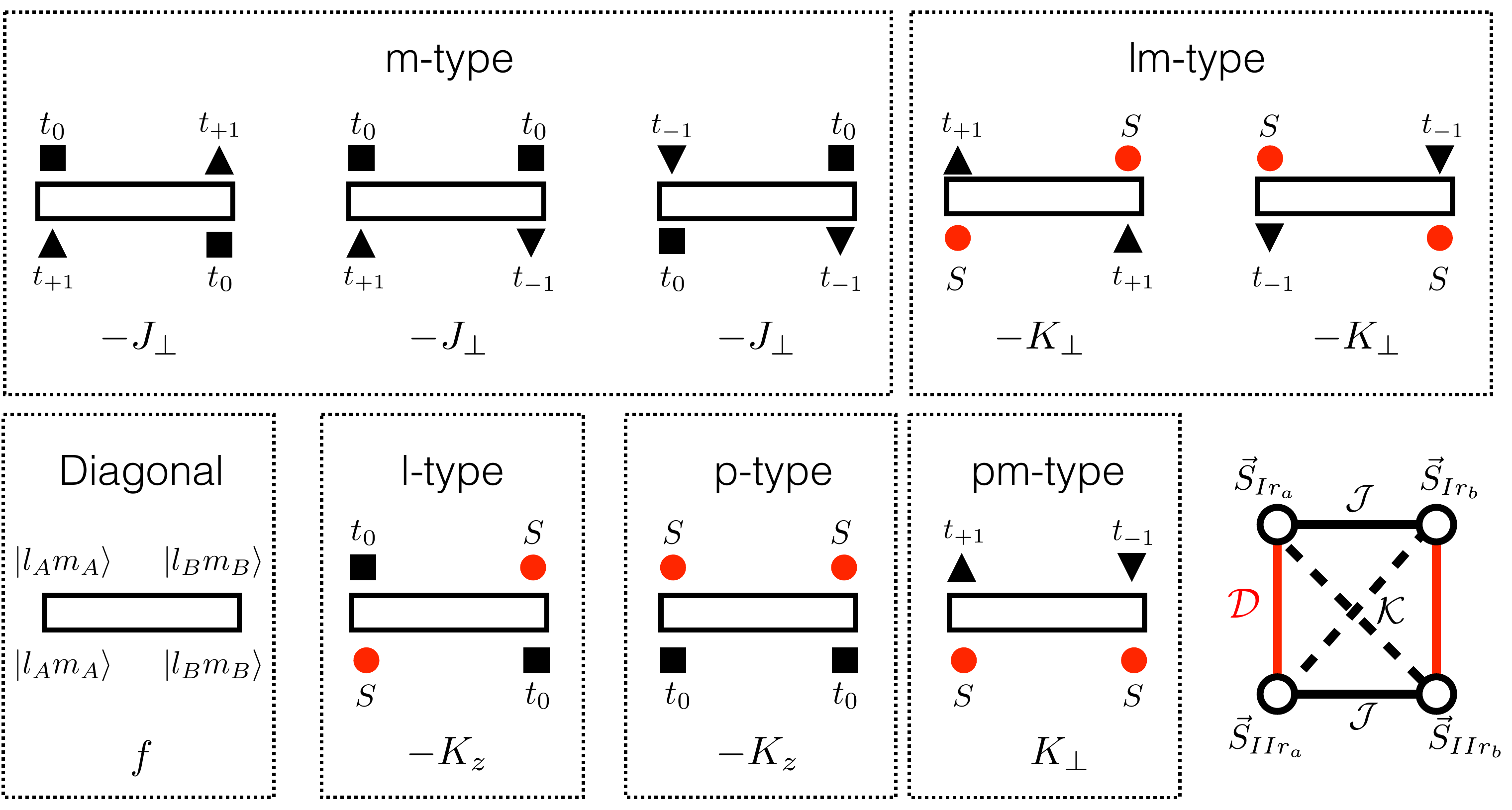}}
\caption{Vertices that appear in the SSE operator
string for $H_{\rm bilayer}$, with corresponding weights in the canonical cluster basis. All other valid vertices are obtained by symmetry operations that
exchange left and right, or upper and lower, legs (keeping the weight fixed). The constant $C$ in the function $f(l_A,l_B,m_A,m_B) = C-J_zm_Am_B-\zeta (\Delta_z-\Delta_\perp)(m_A^2+m_B^2) - \zeta \Delta_\perp (l_A(l_A+1)+l_B(l_B+1))$ is chosen to ensure that $f \geq 0$. Bottom right: Lattice structure and exchange couplings of $H_{\rm bilayer}$. Vertices and lattice structure for $H_{\rm mixed}$ are detailed in the Supplemental Material.}
\label{vertices}
\end{figure}

{\em Synopsis.-} In this Letter,
we introduce a QMC method that works in the basis of spin eigenstates of clusters to simulate a large class of frustrated quantum magnets in a provably sign-free manner. We focus our discussion on systems in which the clusters in question are made
up of two spin $1/2$ moments ($\vec{S}_{{\rm I} r}$ and $\vec{S}_{{\rm II} r}$) located on
{\em layers} ${\rm I}$ and ${\rm II}$ at {\em sites} $r$ of a bipartite Bravais lattice in any spatial dimension (Fig.~\ref{vertices}), with Hamiltonian
\begin{eqnarray}
H_{\rm bilayer} &=& \sum_{\langle r_a r_b \rangle}({\mathcal{J} }_z S^{z}_{{\rm I} r_a} S^{z}_{{\rm I}r_b}+ {\mathcal{J}}_{\perp}\vec{S}^{\perp}_{{\rm I} r_a} \cdot \vec{S}^{\perp}_{{\rm I} r_b} +  {\rm I} \leftrightarrow {\rm II}) + \nonumber \\
&& \sum_{\langle r_a r_b \rangle}({\mathcal {K}}_zS^{z}_{{\rm I}r_a} S^{z}_{{\rm II}r_b}  + {\mathcal{K}}_{\perp} \vec{S}^{\perp}_{{\rm I}r_a} \cdot \vec{S}^{\perp}_{{\rm II}r_b} + {\rm I} \leftrightarrow {\rm II})  +\nonumber \\
&& \sum_{r} ({\mathcal{D}}_{z}S^{z}_{{\rm I}r} S^{z}_{{\rm II}r}+ {\mathcal{D}}_{\perp}\vec{S}^{\perp}_{{\rm I}r} \cdot \vec{S}^{\perp}_{{\rm II}r} ) \; ,
 \end{eqnarray}
where the nearest neighbour links of this Bravais lattice have been denoted by $\langle r_a r_b\rangle$ to emphasize its bipartite nature, and $\vec{S}^{\perp}_{{\rm I}/{\rm II}r}$ represents the vector formed by the two transverse components ($x$ and $y$) of these spins.  Geometric frustration of the exchange interactions leads to a severe sign problem for other QMC methods whenever ${\mathcal D}_{\perp} {\mathcal K}_{\perp} {\mathcal J}_{\perp} > 0$. Our central result is that such frustration leads to no sign problems in our method whenever the interactions in $H_{\rm bilayer}$ are constrained to satisfy {\em at least one} of the following three conditions: i) ${\mathcal{K}}_z = {\mathcal{J}}_z$, ii) ${\mathcal{K}}_{\perp} = {\mathcal{J}}_{\perp}$, iii) ${\mathcal{K}}_{\perp} = -{\mathcal{J}}_{\perp}$. 
{\em Fully-frustrated bilayer} systems~\cite{Gelfand_91,Kolezhuk,Honecker_Mila_Troyer,Brenig_Becker,Chattopadhyay_Bose,Totsuka_Mikeska}, which have infinitely many conserved quantities, represent a special case with i) and ii) both being
satisfied. The method also works when the B-sublattice only hosts a single spin $1/2$ moment $\vec{S}_{r_b}$ that couples symmetrically to $\vec{S}_{{\rm I} r_a}$ and $\vec{S}_{{\rm II} r_a}$ on neighbouring A sublattice sites~\cite{Takano,Niggemann,Honecker_azurite}:
\begin{eqnarray}
H_{\rm mixed} &=&  \sum_{\langle r_a r_b \rangle}({\mathcal{J} }_z S^{z}_{{\rm I} r_a} S^{z}_{r_b}+ {\mathcal{J}}_{\perp}\vec{S}^{\perp}_{{\rm I} r_a} \cdot \vec{S}^{\perp}_{r_b} +  {\rm I} \leftrightarrow {\rm II}) + \nonumber \\
&& \sum_{r_a} ({\mathcal{D}}_{z}S^{z}_{{\rm I}r_a} S^{z}_{{\rm II}r_a}+ {\mathcal{D}}_{\perp}\vec{S}^{\perp}_{{\rm I}r_a} \cdot \vec{S}^{\perp}_{{\rm II}r_a}) \; .
\end{eqnarray}
For ${\mathcal D}_{\perp} > 0$, the usual QMC method has a sign problem, which is no longer present in our QMC scheme.
Our method is also expected to apply to other such models with infinitely many conserved
quantities~\cite{Ivanov,Richter,Koga,Gros,Schulenburg,Honecker_Brenig,Manmana}. Additionally, iii) includes interesting examples of frustrated bilayer magnets with full SU(2) symmetry and no extra conservation laws. Some of our results have been obtained independently in recent parallel work~\cite{Wessel_preprint}.

{\em Key idea.-} We use the Stochastic Series Expansion (SSE) QMC framework~\cite{Sandvik_review} and work at each Bravais lattice site $r$ in the basis $\left \{ |l,m\rangle \right \}$ of simultaneous eigenstates of the total spin $\vec{L}_{r}^2$ and its $z$ component $L^z_{r}$, with eigenvalues $l(l+1)$ and $m$ respectively. For $H_{\rm bilayer}$, we define $\vec{L}_{r} = \vec{S}_{{\rm I} r} + \vec{S}_{{\rm II}r}$ on both sublattices. For $H_{\rm mixed}$, this
is modified on the $B$ sublattice by defining $\vec{L}_{r_b} = \vec{S}_{r_b}$. We decompose the Hamiltonian into terms 
living on bonds $\langle r_a r_b \rangle$ of the bipartite Bravais lattice, with the terms proportional to ${\mathcal{D}}_z$ and ${\mathcal{D}}_\perp$ at each site $r$ being shared equally among all bonds emanating from $r$:
\begin{eqnarray}
&H_{1\langle r_a r_b\rangle} &= J_zL^{z}_{r_a}L^{z}_{r_b} + \zeta \Delta_{z}\left[(L^{z}_{r_a})^2+(L^{z}_{r_b})^2\right] \nonumber \\
&&+ \zeta \Delta_{\perp}\left[( \vec{L}^{\perp}_{r_a} )^2 + (\vec{L}^{\perp}_{r_b})^2\right] -C \; ,\nonumber \\
&H_{2\langle r_a r_b\rangle}^{\pm} &= \frac{J_{\perp}}{2} ( L^{\pm}_{r_a} \cdot L^{\mp}_{r_b})\; ,\nonumber \\ 
& H_{3\langle r_a r_b\rangle}&=K_zN^{z}_{r_a}N^{z}_{r_b} \; ,\nonumber\\
&H_{4\langle r_a r_b\rangle}^{\pm}&=\frac{K_{\perp}}{2} (N^{\pm}_{r_a} \cdot N^{\mp}_{r_b}) \;,
\label{bilayerdecomposition}
\end{eqnarray}
with $\vec{L}^{\perp}_r$ the vector made of transverse ($x/y$) components of
$\vec{L}_r$, $\vec{N}_{r} = \vec{S}_{{\rm I}r} - \vec{S}_{{\rm II}r}$, $L^{\pm}_r = L^x_r \pm i L^y_r$,
$N^{\pm}_r=N^x_r \pm i N^y_r$, $\zeta$ the
inverse coordination number of the bipartite lattice, $C$ a constant
introduced to ensure negativity of all matrix elements of the {\em diagonal} operator $H_{1\langle r_a r_b\rangle}$ in our chosen basis, and $\Delta_\mu = {\mathcal D}_\mu/2$, $J_{\mu}=({\mathcal J}_\mu+{\mathcal K}_\mu)/2$, $K_\mu=({\mathcal J}_\mu-{\mathcal K}_\mu)/2$ (for $\mu = z, \perp$). Using this decomposition, we have
$H_{\rm bilayer}= \sum_{\langle r_a r_b \rangle} H_{1\langle r_a r_b\rangle} + H_{3\langle r_a r_b\rangle} + H_{2\langle r_a r_b\rangle}^{+}  +  H_{4\langle r_a r_b\rangle}^{+}+ H_{2\langle r_a r_b\rangle}^{-} +  H_{4\langle r_a r_b\rangle}^{-}$. $H_{\rm mixed}$, when decomposed in the same way, only has analogs of the $H_{1\langle r_a r_b \rangle}$ and $H_{2\langle r_a r_b \rangle}^{\pm}$ terms (Supplemental Material). 

Working within the SSE framework with this decomposition of $H_{\rm bilayer}$, one writes 
\begin{eqnarray}
Z\! \! &=& \! \! \! \!  \! \sum_{n=0}^{\infty} \frac{1}{n!T^n}\sum_{{\mathcal S}_n} \langle \alpha_0 | (-H_n)|\alpha_{n-1}\rangle \langle \alpha_{n-1}| (-H_{n-1})|\alpha_{n-2}\rangle\dots \nonumber \\
&&\!\!\!\!\!\!\dots \langle \alpha_1| (-H_1)|\alpha_0\rangle \;, \!\!\!\!\!\!\!\!\!\!\!\!\!\!\!\!\!
\end{eqnarray}
where the sum over {\em operator-strings} ${\mathcal S}_n$ of length $n$ is implemented
by allowing each $|\alpha_j\rangle$ to range over the full basis of states,
and each $H_j$ to range over all bond-operators $H_{1\langle r_a r_b\rangle}$,
$H_{2\langle r_a r_b\rangle}^{\pm}$,  $H_{3\langle r_a r_b\rangle}$, and $H_{4\langle r_a r_b\rangle}^{\pm}$.  Along with the factor of $1/(n!T^n)$, the product of matrix elements appearing in the summand serves as the Monte Carlo weight of each operator-string, and the QMC simulation proceeds
by performing an importance sampling of $Z$.

{\em Proof of positive-weight property.-} Clearly, the weight of an operator string {\em does not}
depend on the choice of arbitrary phase factors attached to individual basis states, since
these phase factors cancel in pairs in the product of matrix elements that sets the
weight. Fixing these phases, we define the {\em canonical} cluster
basis as follows: $|s\rangle = ( |\!\!\uparrow_{{\rm I}} \downarrow_{{\rm II}}\rangle - |\!\!\downarrow_{{\rm I}} \uparrow_{{\rm II}}\rangle)/\sqrt{2}$, $|t_{+1}\rangle =  |\!\!\uparrow_{{\rm I}} \uparrow_{{\rm II}}\rangle$, $|t_{-1}\rangle =  |\!\!\downarrow_{{\rm I}} \downarrow_{{\rm II}}\rangle$, and
$|t_0\rangle = (|\!\! \uparrow_{{\rm I}} \downarrow_{{\rm II}}\rangle + |\!\! \downarrow_{{\rm I}} \uparrow_{{\rm II}}\rangle)/\sqrt{2}$. Next, we classify the off-diagonal matrix elements contributing to the weight of an operator-string into five types (Fig.~\ref{vertices}):  (i) {\em $m$-type processes} that hop one quantum of $L^z$ along link $\langle r_a r_b \rangle$ between
two neighbouring triplet clusters, (ii) {\em $l$-type processes} that exchange the states $|s\rangle$ and $|t_0\rangle$ of two neighbouring clusters, (iii) {\em $p$-type processes} that take neighbouring singlet clusters  and promote both to the $|t_0\rangle$ state or vice-versa, (iv) {\em $lm$-type processes} that exchange singlet and triplet states of neighbouring clusters and simultaneously hop one quantum of $L^z$,
and (v) {\em $pm$-type processes} that take neighbouring singlet clusters 
to states $|t_{\pm1}\rangle$ and $|t_{\mp1}\rangle$ respectively, or vice versa.  

All processes of a given type have a fixed sign for the corresponding matrix elements between basis states (Fig.~\ref{vertices}). 
Therefore, a positive
weight is guaranteed if ${\mathcal N}_t$, the number of occurrences (in any string ${\mathcal S}_n$) of $t$-type processes, has even parity for each type $t$. These parities are constrained by the periodicity of the operator-string ${\mathcal S}_n$, {\em i.e.} the starting state $|\alpha_0\rangle$ is recovered after the action of $n$ operators. Since pair creation of the $l$ quantum number must be balanced by
corresponding pair destruction processes, ${\mathcal N}_p+{\mathcal N}_{pm}$ must be even. Since a bipartite lattice only has loops of even length, the number of occurrences of processes that hop the $m$ quantum number must be even, implying that ${\mathcal N}_m+ {\mathcal N}_{pm} + {\mathcal N}_{lm} $ is even. By the same argument applied to the $l$ quantum number, ${\mathcal N}_l+{\mathcal N}_{lm} $ must also be even.

Since $H_{2 \langle r_a r_b \rangle}^{\pm}$ only gives rise
to $m$-type processes, while $H_{3 \langle r_a r_b \rangle}$ gives rise
to $l$-type and $p$-type processes and $H_{4 \langle r_a r_b \rangle}^{\pm}$
gives rise to $lm$-type and $pm$-type processes,  $K_{\perp} = 0$ ($K_z=0$) implies ${\mathcal N}_{lm} = {\mathcal N}_{pm}=0$ (${\mathcal N}_l={\mathcal N}_p=0$). The periodicity constraints
then imply that all nonzero ${\mathcal N}_{t}$ are even in both these cases. As a result, in both these
cases, each ${\mathcal S}_n$ has positive weight in this QMC scheme {\em regardless
of the sign of} all nonzero couplings. On the other hand, if $J_\perp=0$,  {\em i.e.} ${\mathcal N}_m=0$, only $pm$-type processes can create or destroy pairs of $m=\pm 1$ states on neighbouring sites, thus ensuring that ${\mathcal N}_{pm}$ is even. Along with the other periodicity constraints, this again implies
that all nonzero ${\mathcal N}_t$ are even, yielding a sign-problem-free method whenever $J_\perp=0$, {\em independent of the sign of} other couplings.

Thus, the Monte Carlo weight is positive for frustrated bilayer magnets with Hamiltonian $H_{\rm bilayer}$ when {\em at least one} of the following conditions is satisfied: i) ${\mathcal{K}}_z = {\mathcal{J}}_z$, ii) ${\mathcal{K}}_{\perp} = {\mathcal{J}}_{\perp}$, iii) ${\mathcal{K}}_{\perp} = -{\mathcal{J}}_{\perp}$.  When i) and ii) are both satisfied, one obtains fully frustrated bilayer systems~\cite{Gelfand_91,Kolezhuk,Honecker_Mila_Troyer,Brenig_Becker,Chattopadhyay_Bose,Totsuka_Mikeska} with infinitely many conservation laws, which can have either SU(2) or
U(1) symmetry. Additionally, iii) also contains other examples of SU(2) symmetric frustrated magnets (e.g. for $J_z=J_\perp=0$, $K_z=K_\perp >0$ and $\Delta_z=\Delta_\perp < 0$) and no extra conservation laws. A similar argument establishes the absence of a sign problem for $H_{\rm mixed}$~(Supplemental Material).

Alternately, this positive-weight property can be made explicit by switching from the canonical cluster basis to a rotated basis obtained by attaching phase-factors $e^{i\theta_{|lm\rangle}+i\eta_{|lm\rangle}}$ to the states $|l,m\rangle$. Here, the $\theta_{|lm\rangle}$ are $r$-independent, while the $\eta_{|l,m\rangle}$ are $0$ on the $B$ sublattice and constant on the $A$ sublattice. These phases are chosen in each of these
three generic cases to ensure that every nonzero contribution to the weight in the rotated basis is explicitly positive. For instance, when $K_{\perp}=0$, we set $\eta_{|t_{\pm}\rangle}=\theta_{|l,m\rangle}=0$ (for all $l$, $m$), while independently choosing $\eta_{|s\rangle}$ to be $0$ or $\pi$
and $\eta_{|t_0\rangle}$ to be $0$ or $\pi$ depending on the signs of $K_z$ and $J_\perp$.
The other sign-free cases can be handled with slightly different choices for these phase factors~(Supplemental Material).  The positive-weight property of $H_{\rm mixed}$ can also be made explicit in the same way~(Supplemental Material).

{\em Implementation.-} A key advantage of this QMC approach is that the usual SSE framework~\cite{Sandvik_review} remains valid with no change in the construction of the diagonal update, and one new feature in the construction of directed loop updates: {\em Three} different kinds of directed loop updates~\cite{syljuasen,alet} are now possible, involving changes to the $m$ quantum number, or the $l$ quantum number, or both.
Additionally, to improve statistics, one can use parallel tempering~\cite{hukushima} 
as well as an additional local update, which identifies worldlines that are only touched by diagonal vertices, and changes their state using Metropolis-type acceptance probabilities. 
\begin{figure}
{\includegraphics[width=\columnwidth]{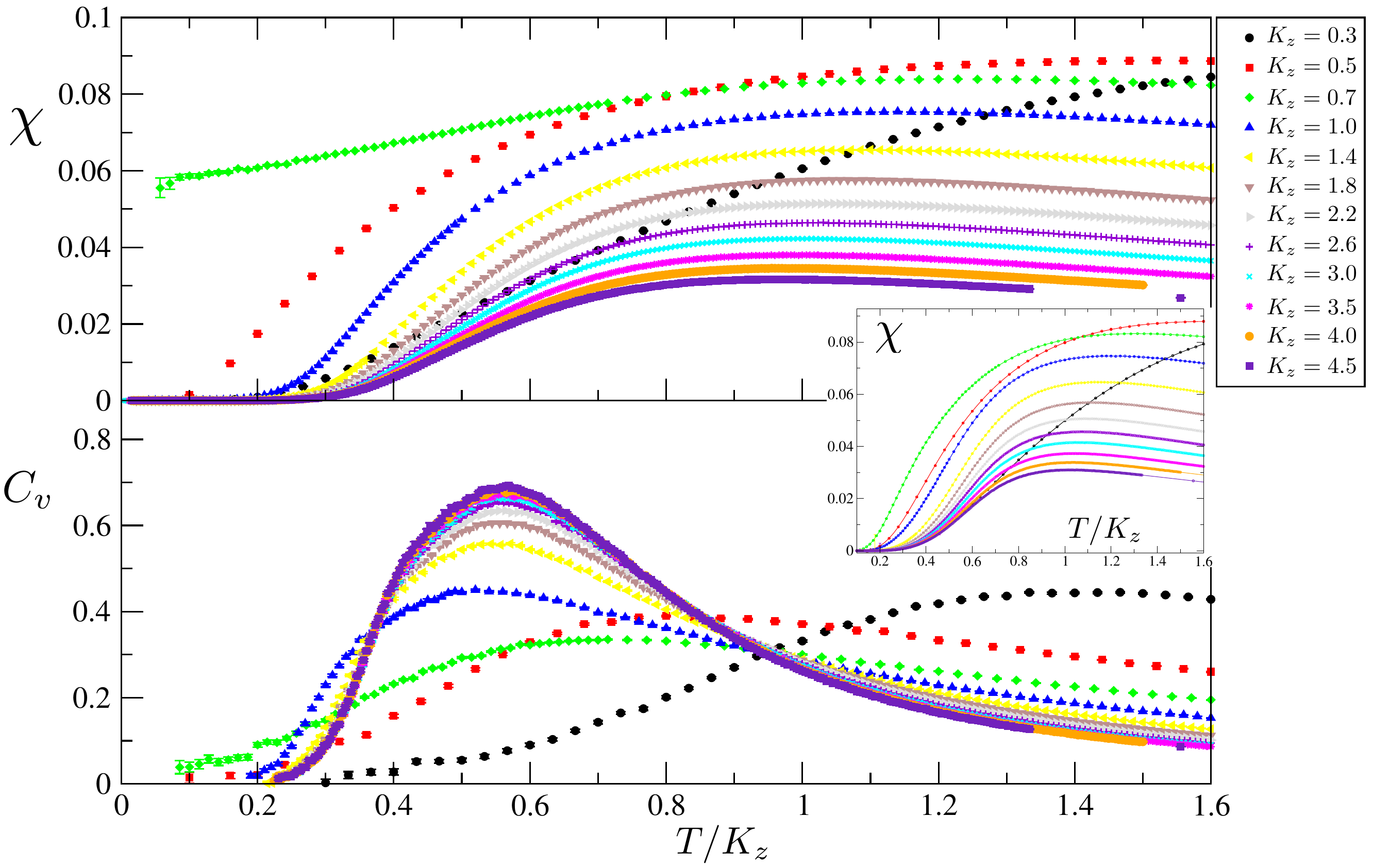}}
\caption{Temperature ($T$) dependence of the susceptibility and specific heat of of $H_{\rm bilayer}$  with ${\cal D}_z={\cal D}_\perp=1$, ${\mathcal J}_\perp = {\mathcal K}_\perp = 1$, ${\mathcal J}_z = 1+K_z$, ${\mathcal K}_z = 1-K_z$. Symbols display data for a sample with $L=64$ unit cells, plotted for a variety of values of $K_z$. The inset shows the perfect agreement between QMC data (symbols) and exact diagonalization results (lines) for $L=6$.}
\label{Kappa}
\end{figure}

{\em Benchmarks.-} Our implementation, which incorporates all these updates, has been
successfully benchmarked against exact diagonalization in spatial dimension $d=1$ for all the sign-free cases, including the two special cases with infinitely many conserved quantities~(Supplemental Material).
In Fig.~\ref{Kappa} (inset), we illustrate this for a representative example, focusing on the susceptibility $\chi=\beta\langle (S^z)^2 \rangle/N$  and specific heat per spin $C_v=\frac{1}{N}\frac{d \langle H \rangle}{dT}$ for $H_{\rm bilayer}$ in $d=1$, with $K_\perp=0$ ($N=2L$ is the number of spins $1/2$ in a ladder of length $L$). Data in the main panel illustrate the power of the method, which allows us to access the thermodynamics of this frustrated ladder for fairly large $L$ up to low T, and for a range of values of $K_z$.
 
\begin{figure*}[t]
{\includegraphics[width=1.8\columnwidth]{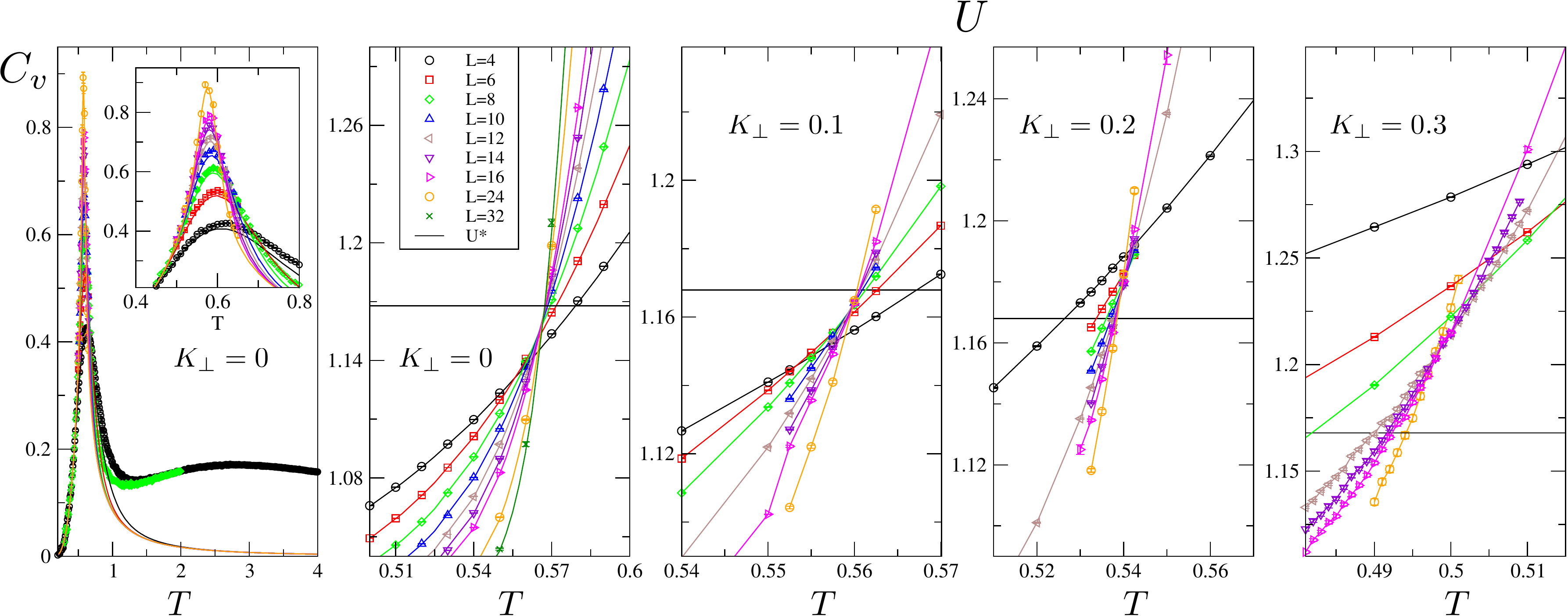}}
\caption{QMC results (symbols) for $H_{\rm bilayer}$ in a field on the square lattice, with ${\cal J}_{z}={\cal K}_{z}=1$, ${\cal D}_z={\cal D}_{\bot}=5$, ${\mathcal J}_\perp = 1+K_\perp$, ${\mathcal K}_\perp = 1-K_\perp$, and $h=7$. Left panel: Specific heat $C_v$ for $K_\perp=0$ (inset zooms into the critical range). Right panels: Binder cumulant $U=\langle m_s^4 \rangle / \langle m_s^2 \rangle^2$ of the staggered magnetization $m_s=\sum_r (-)^{r} (S^z_{Ir}+S^z_{IIr})$. The critical temperature $T_c$, estimated by the crossing point of $U$, decreases with $K_\perp$. $U$ at the estimated $T_c$ tends to the 2d Ising critical value $U^*=1.16793$~\cite{Blote} at large $L$ for all $K_\perp$ displayed. The solid lines in the $K_{\perp} >0$ panels are guides to the eye. At $K_{\perp}=0$, they denote results for the 2d classical Ising model at $T_{\rm Ising}=4T$.}
\label{fig:2d}
\end{figure*}

{\em Numerical results in d=2.-} We now consider $H_{\rm bilayer}$ on a square Bravais lattice in the presence of an additional {\it magnetic field}, which our method can handle without a sign problem: $H=H_{\rm bilayer}-h\sum_r (S^z_{Ir}+S^z_{IIr})$ (the magnetic field only modifies weights of the diagonal vertices). The physics of the fully-frustrated special case ($K_z = K_\perp = 0$) in a certain field range
was argued~\cite{Richter_Derzhko} to be dominated at low temperature by sublattice-ordered configurations in which two-spin clusters on one spontaneously chosen sublattice are
in the triplet state $| t_{+1} \rangle$, while two-spin clusters on the other sublattice   are in a singlet state $|s\rangle$, allowing a low-temperature mapping to the ordered phase of hard squares on the
square lattice. This predicts that the system undergoes a temperature-driven phase transition in the 2d Ising universality class to a high-temperature phase in which sublattice symmetry is restored~\cite{Richter_Derzhko,Derzhko}. 

We have performed a high-precision QMC test of this prediction 
using a finite-size scaling analysis for samples with up to $N=2L^2=2048$ spins $1/2$, both for the fully-frustrated special case, and in the presence of nonzero $K_\perp$ (previous quantum simulations were limited to $K_\perp=0$ and $N=20$). Our determination of the critical value of the Binder cumulant of the staggered magnetization
provides clear evidence that this transition indeed belongs to the 2d Ising universality class both for $K_\perp=0$ {\em and} for nonzero $K_\perp$ (albeit with stronger finite-size effects in this case). Indeed, our QMC data for the specific heat and the Binder cumulant close to the phase transition (Fig.~\ref{fig:2d}) are almost identical to those of the classical 2d Ising model when $K_\perp=0$, but deviate from the classical results outside the critical region, underscoring the nontrivial nature of this correspondence. These deviations become much more significant for nonzero $K_\perp$. Our method thus enables an investigation of the full parameter regime, including where the hard-square mapping breaks down, both in the fully-frustrated
special case and when $K_\perp \neq 0$ (the effect of $K_z \neq 0$
can also be studied).

\begin{figure}
{\includegraphics[width=0.85\columnwidth]{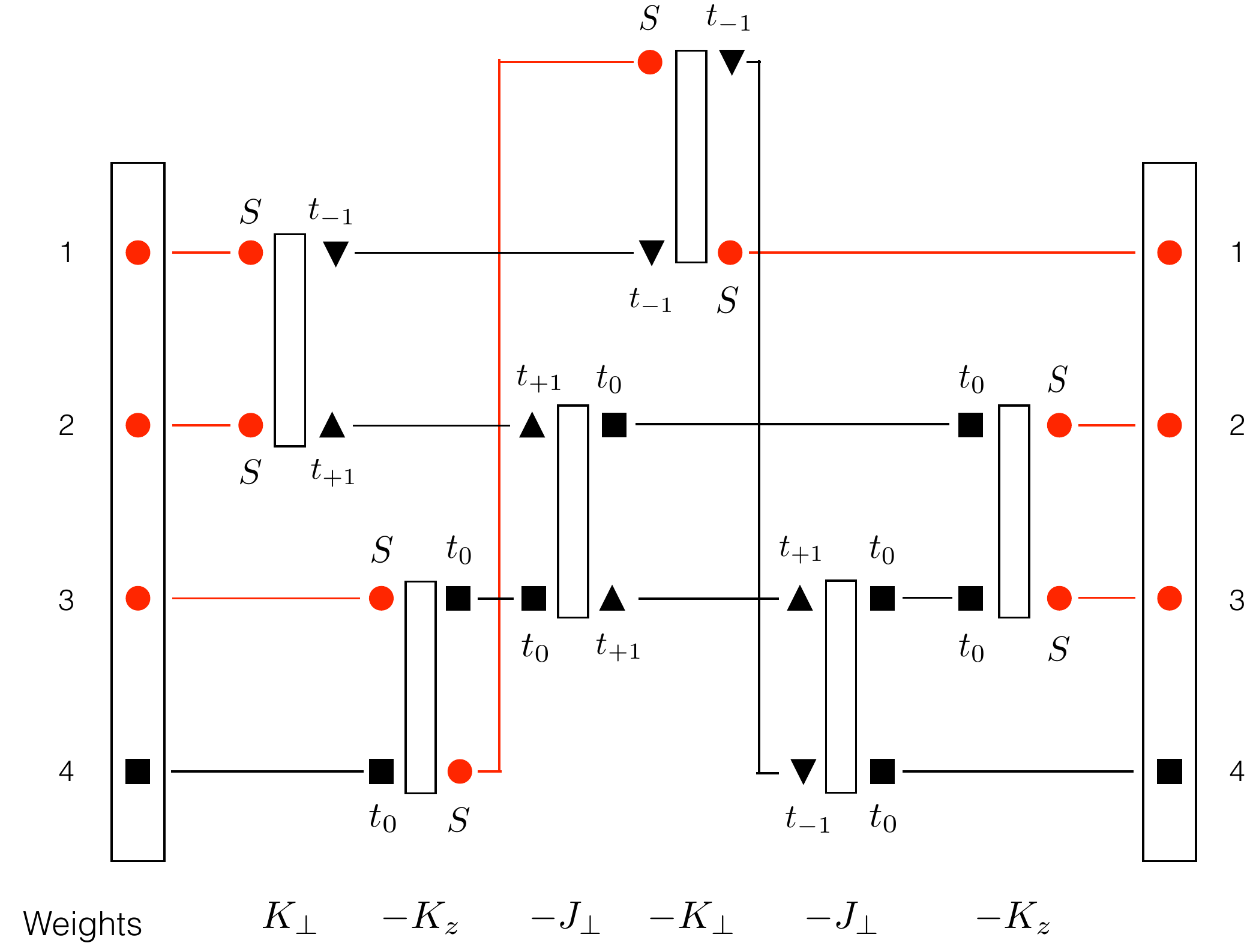}}
\caption{This operator string for a single plaquette of the Bravais lattice
of the bilayer system illustrates the origin of the sign problem faced when
simulating the general bilayer Hamiltonian $H_{\rm bilayer}$: its weight is negative, independent of the signs of the nonzero couplings $J_{\perp}$, $K_z$, and $K_{\perp}$.}
\label{signproblem}
\end{figure}

{\em Discussion.- } Clearly, the method presented here can be applied to a large class of frustrated magnets~\cite{Muller,Lin,Richter_Derzhko,Rossler}, and models closely related to specific strongly correlated materials: for instance, a generalized version of $H_{\rm mixed}$ has been argued to be a good model for the mineral azurite  Cu$_3$(CO$_3$)$_2$(OH)$_2$~\cite{Honecker_azurite}, and the specific heat of the fully frustrated ladder~(Supplemental Material) has similar features with the Shastry-Sutherland compound SrCu$_2$(BO$_3$)$_2$~\cite{Wessel_preprint}. This QMC method also enables the search for finite-T signatures of multi-triplet bond states, as  shown in Ref.~\cite{Wessel_preprint}. Additionally, it offers the possibility of using large-scale unbiased simulations to study interesting quantum phase transitions driven by the competition between different exchange interactions. As we illustrated, a magnetic field (in the $z$ direction) can also be
included, thus allowing one to study magnetization processes and plateaux of such frustrated magnets~\cite{Honecker_Mila_Troyer,Schulenburg,Totsuka_Mikeska,Manmana}. 
On the flip-side, we note that this QMC scheme does {\em not} remain sign-free when $J_\perp$, $K_{\perp}$ and $K_{z}$ are {\em all} nonzero in the general bilayer Hamiltonian $H_{\rm bilayer}$. The simple sequence of processes shown in Fig.~\ref{signproblem} for a single plaquette of
a square lattice provides an explicit illustration of this limitation. Nevertheless, this construction of negative-weight configurations {\em relies on the existence of loops} in the underlying bipartite Bravais lattice, and leaves open the possibility that this sign problem could be controlled in 1d systems with open boundaries. To summarize, our work has led to a solution of the sign problem for a large and interesting class of frustrated quantum magnets. Given the ubiquity of the sign problem in computational physics, we hope that the strategy outlined in this work can be adapted to other systems.

{\em Acknowledgments.-} We would like to thank S. Wessel for generously sharing with us some
details of the recent parallel work Ref.~\cite{Wessel_preprint}, and for stimulating discussions regarding the sign problem in the general case. We gratefully acknowledge J. Richter for suggesting the study of the finite-temperature Ising transition in the square bilayer system in a field.
One of us (KD) would also like to gratefully acknowledge the hospitality of ECT* (Trento), ICTP (Trieste), and ICTS-TIFR (Bengaluru) during the writing of this manuscript.
This work was  supported by the Indo-French Centre for the Promotion of Advanced Research (IFCPAR/CEFIPRA Project 4504-1), and performed using numerical resources from GENCI (grants 2014-x2014050225 and 2015-x2015050225) and CALMIP. Our QMC code is based on the SSE code of the ALPS libraries~\cite{SSE,ALPS13}.

\newpage
\onecolumngrid
\appendix

\section{Supplemental material for ``Sign-problem-free Monte Carlo simulation of certain frustrated quantum magnets''}

\subsection{Decomposition of $H_{\rm mixed}$ and SSE vertices}

In order to decompose $H_{\rm mixed}$ into operators that live on links $\langle r_a r_b \rangle$ of the underlying bipartite Bravais lattice, we define
\begin{eqnarray}
&H_{1\langle r_a r_b\rangle} &= J_zL^{z}_{r_a}L^{z}_{r_b} + \zeta \Delta_{z}\left[(L^{z}_{r_a})^2\right] \nonumber \\
&&+ \zeta \Delta_{\perp}\left[( \vec{L}^{\perp}_{r_a} )^2 \right] -C_{\rm mixed} \; ,\nonumber \\
&H_{2\langle r_a r_b\rangle}^{\pm} &= \frac{J_{\perp}}{2} ( L^{\pm}_{r_a} \cdot L^{\mp}_{r_b})\; ,
\label{mixeddecomposition}
\end{eqnarray}
where $\vec{L}^{\perp}_{r_a}$ denotes the vector made of transverse ($x/y$) components of $\vec{L}_{r_a} \equiv \vec{S}_{{\rm I} r_a} + \vec{S}_{{\rm II} r_b}$ and $\vec{L}^{\perp}_{r_b}$ denotes the vector made of transverse ($x/y$) components of $\vec{L}_{r_b} \equiv \vec{S}_{r_b}$, $L^{\pm}_r = L^x_r \pm i L^y_r$, $\zeta$ is the inverse coordination number of the bipartite lattice, $C_{\rm mixed}$ is a constant introduced to ensure that all matrix elements of the {\em diagonal} operator $H_{1\langle r_a r_b\rangle}$ are negative in our chosen basis, $\Delta_\mu = {\mathcal D}_\mu/2$,  and $J_{\mu}={\mathcal J}_\mu$ (for $\mu = z, \perp$). Using this decomposition, we have $H_{\rm mixed}= \sum_{\langle r_a r_b \rangle} H_{1\langle r_a r_b\rangle} + H_{2\langle r_a r_b\rangle}^{+}  +H_{2\langle r_a r_b\rangle}^{-}$.

\subsection{Details for Quantum Monte Carlo simulations of $H_{\rm mixed}$}

The QMC simulations for $H_{\rm mixed}$ use this decomposition to work
within the Stochastic Series Expansion using diagonal updates and directed
loop updates that change the $m$ quantum number locally during the
construction of the loop.

\
\begin{figure}[h]
{\includegraphics[width=0.5\columnwidth]{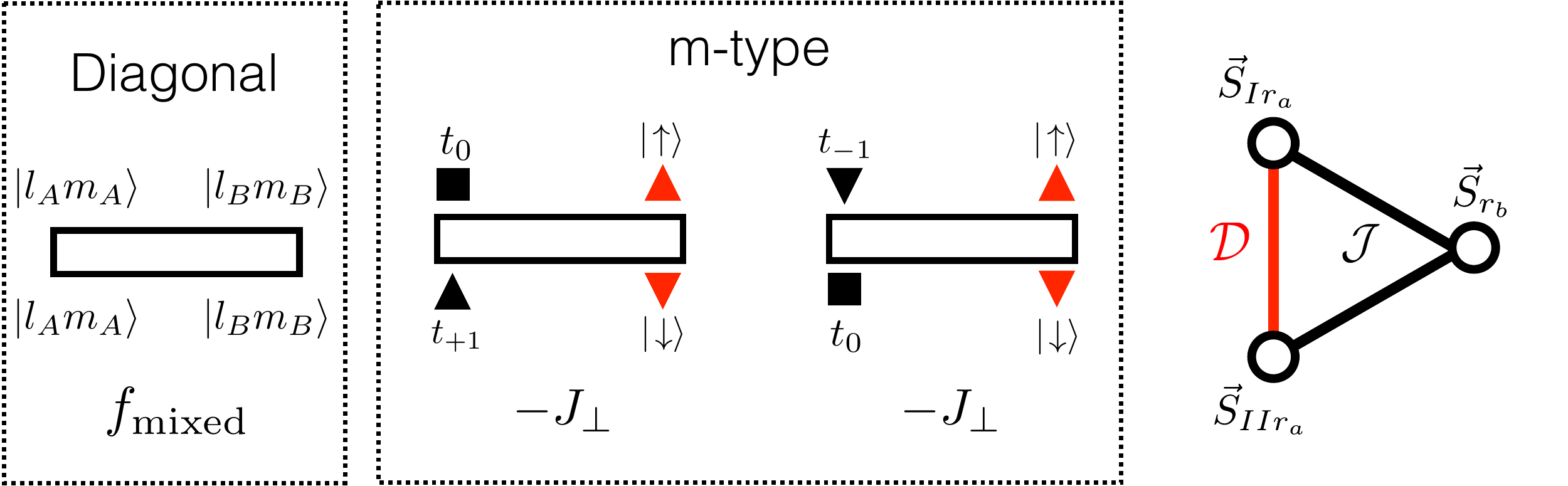}}
\caption{Vertices that appear in the SSE operator
string for $H_{\rm mixed}$, with corresponding weights in the canonical cluster basis. Other valid vertices are obtained by symmetry operations that
exchange upper and lower legs (keeping the weight fixed). The constant $C_{\rm mixed}$ in the function $f_{\rm mixed} (l_A,l_B,m_A,m_B) = C_{\rm mixed} - J_zm_Am_B-\zeta (\Delta_z-\Delta_\perp) m_A^2- \zeta \Delta_\perp l_A(l_A+1)$ is chosen to ensure that $f_{\rm mixed}  \geq 0$. Right panel: Pictorial representation of the lattice structure and
exchange interactions of $H_{\rm mixed}$.}
\label{vertices_mixed}
\end{figure}

\subsection{Explicit demonstration of positive-weight property}

For $H_{\rm bilayer}$, the positive-weight property of our method can be made explicit by switching from the canonical cluster basis to a rotated basis obtained by attaching phase-factors $e^{i\theta_{|lm\rangle}+i\eta_{|lm\rangle}}$ to the states $|l,m\rangle$ of the canonical basis. Here, the $\theta_{|lm\rangle}$ are completely $\vec{r}$-independent, while the $\eta_{|l,m\rangle}$ are $0$ on the $B$ sublattice and constant on the $A$ sublattice. These phases are chosen in each of these three generic cases to ensure that every nonzero contribution to the weight in the rotated basis is explicitly positive. For instance, as already mentioned in the main text, when $K_{\perp}=0$, we set $\eta_{|t_{\pm}\rangle}=\theta_{|l,m\rangle}=0$ (for all $l$, $m$), while independently choosing $\eta_{|s\rangle}$ to be $0$ or $\pi$ and $\eta_{|t_0\rangle}$ to be $0$ or $\pi$ depending on the signs of $K_z$ and $J_\perp$. When $K_{z}=0$,  we set $\eta_{|t_{\pm}\rangle}=\theta_{|s\rangle}=0$, and, depending on the signs of $K_\perp$ and $J_{\perp}$, choose $\theta_{|t_0\rangle}=\theta_{|t_{\pm}\rangle}$  to be either $0$ or $\pi/2$, while independently choosing $\eta_{|s\rangle}$ to be $0$ or $\pi$ and $\eta_{|t_0\rangle}$ to be $0$ or $\pi$. Finally, when $J_\perp=0$, we set $\eta_{|t_{\pm}\rangle}=\theta_{|t_0\rangle}=\theta_{|s\rangle}=0$, and, depending on the signs of $K_\perp$ and $K_z$, choose $\theta_{|t_{+}\rangle}=\theta_{|t_{-}\rangle}$  to be either $0$ or $\pi/2$, while independently choosing $\eta_{|s\rangle}$ to be $0$ or $\pi$ and $\eta_{|t_0\rangle}$ to be $0$ or $\pi$. 

For $H_{\rm mixed}$, we see from Fig.~\ref{vertices_mixed} that the only off-diagonal vertices are $m$-type vertices (in the terminology used in the main text). Now, as already noted in the main text, the fact that a bipartite lattice only has loops of even length implies that the number of occurrences of processes that hop the $m$ quantum number must be even in order to satisfy the requirement of periodicity of the operator string. For $H_{\rm mixed}$, this implies that ${\mathcal N}_m$ be even. This establishes the positive-weight property of our method for $H_{\rm mixed}$, since any minus signs arising from off-diagonal operators come in pairs. This can be made explicit exactly as in the discussion above for $H_{\rm bilayer}$, simply by choosing a rotated basis obtained by attaching a phase factor $\eta_{|t_0\rangle}$ that is chosen to be equal to $0$ or $\pi$ depending on the sign of $J_{\perp}$ (all other $\eta$ and $\theta$ are set to zero).

\subsection{Benchmark results on $d=1$ systems}

Here we present results for two one-dimensional systems that serve as our benchmarks. These are the so-called fully-frustrated ladder and the diamond chain~\cite{Honecker_Mila_Troyer,Honecker_azurite}. The former represents the one-dimensional special case of $H_{\rm bilayer}$ with $K_z=K_{\perp}=0$, while the latter is the one-dimensional realization of $H_{\rm mixed}$. These results on the benchmarks, together with benchmarking done in the main text for $H_{\rm bilayer}$ in one dimension with $K_{z} \neq 0$, illustrate that our new QMC method  clearly  allows  us  to  reliably  compute  the  thermodynamics  of  large  systems  over  a  wide  temperature range.

\subsubsection{Fully frustrated ladder}

\begin{figure}[ht]
{\includegraphics[width=0.5\columnwidth]{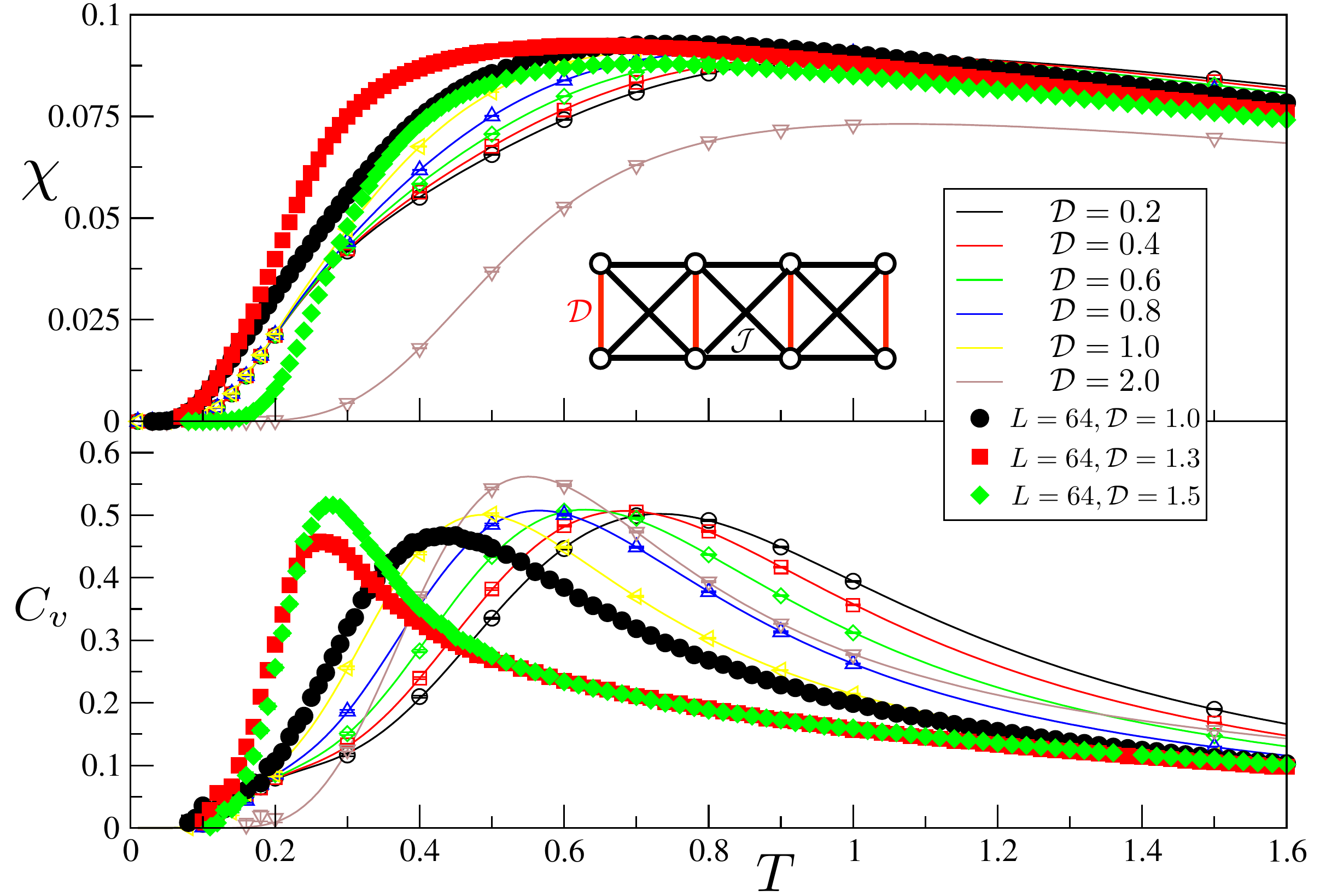}}
\caption{Susceptibility and specific heat versus temperature for the fully-frustrated ladder, {\it i.e.} the $d=1$ bilayer Hamiltonian $H_{\rm bilayer}$ with ${\cal J}_{\perp}={\cal K}_\perp=1$, ${\cal J}_{z}={\cal K}_z=1$, and various values of ${\cal D}_z={\cal D}_\perp \equiv {\cal D}$. Solid lines are exact diagonalization results for a small system with $L=6$, and the corresponding QMC results are depicted by open symbols. Filled symbols are QMC data for a larger ladder of linear size (number of unit-cells) $L=64$.}
\label{fig:FF}
\end{figure}

As mentioned above, this corresponds to a one-dimensional special case of $H_{\rm bilayer}$ where ${\cal K}_z={\cal J}_z$ {\it and} ${\cal K}_\perp={\cal J}_{\perp}$ corresponding to the conditions i) and ii) of the main text being both satisfied. In the example of Fig.~\ref{fig:FF}, we set ${\cal J}={\cal J}_z={\cal J}_\perp={\cal K}_z={\cal K}_{\perp}=1$ and vary ${\cal D}={\cal D}_z={\cal D}_\perp$ (this model has SU(2) symmetry). The lattice and notations are depicted in the top panel of Fig.~\ref{fig:FF} for completeness. This figure display QMC results (symbols) obtained with the algorithm described in the main text for the specific heat $C_v$ and magnetic susceptibility $\chi$ for ladders of linear size $L=6$ and $L=64$ (containing $2L$ spins 1/2) with periodic boundary conditions. The $L=6$ data are displayed to show the perfect agreement with exact diagonalization data for $L=6$.

\subsubsection{Diamond chain}

\begin{figure}[ht]
{\includegraphics[width=0.5\columnwidth]{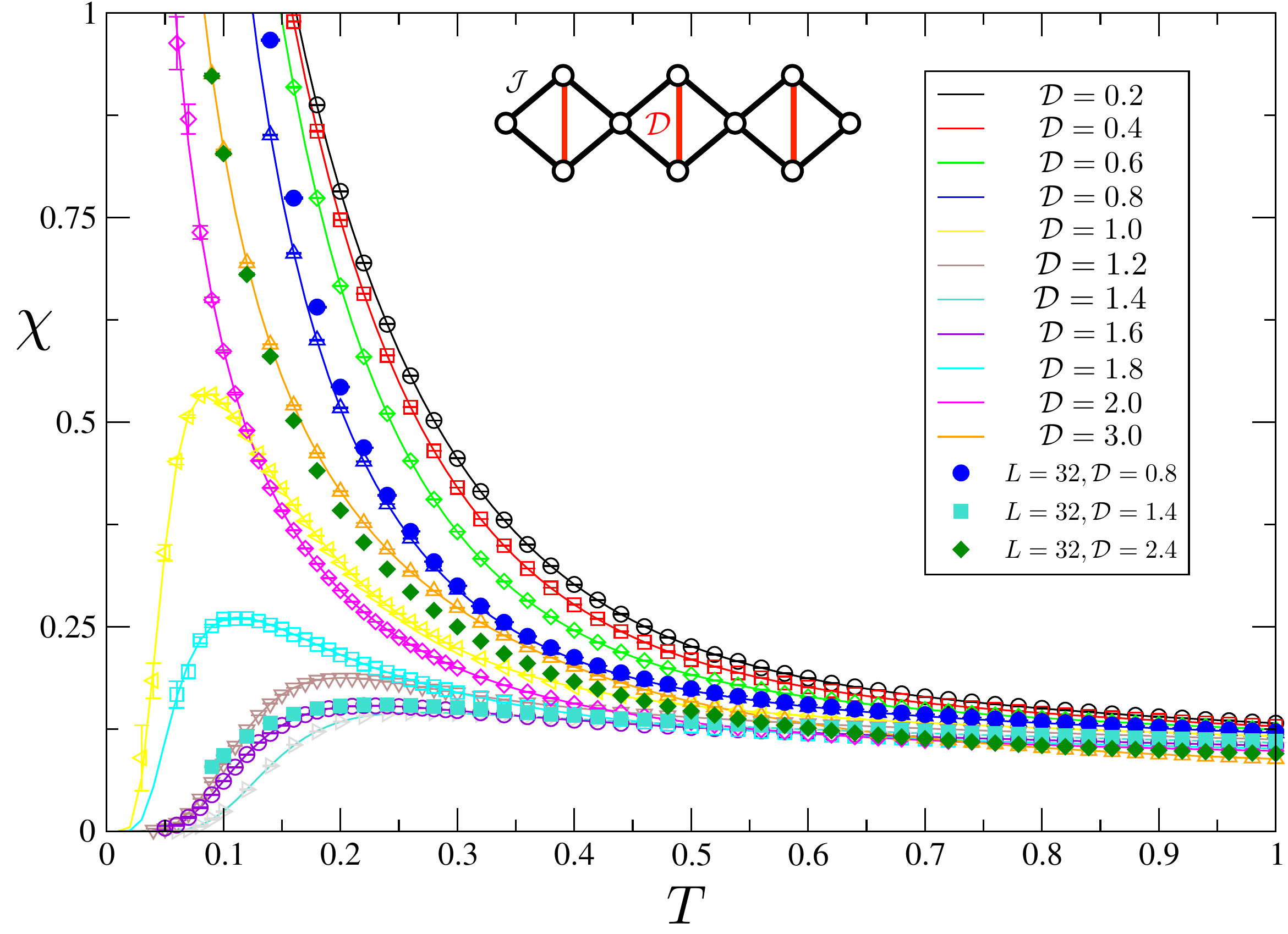}}
\caption{Susceptibility versus temperature for the special case of $H_{\rm mixed}$ in $d=1$ (``diamond chain''), with ${\cal J}_{\perp}={\cal J}_z=1$  for various values of ${\cal D}_z={\cal D}_\perp \equiv {\cal D}$. Solid lines are exact diagonalization results  for a small system with $L=4$ diamond unit cells ({\it i.e.} $N=12$ spins), with corresponding QMC data depicted by open symbols. Filled symbols are QMC data for a larger diamond chain with $L=32$ unit cells.}
\label{fig:diamond}
\end{figure}

As mentioned above, the diamond chain is a one-dimensional realization of $H_{\rm mixed}$. Here, we consider the SU(2) symmetric case, setting ${\cal J}={\cal J}_z={\cal J}_\perp=1$ and varying ${\cal D}={\cal D}_z={\cal D}_\perp$. The lattice is represented in the inset of Fig.~\ref{fig:diamond} for completeness, and contains three spin-half moments per unit cell. Fig.~\ref{fig:diamond} represents the temperature dependence of the magnetic susceptibility for various values of ${\cal D}$ for chains with $L=4$ and $L=32$ units cells ($N=12$ and $N=96$ spins 1/2 in total), as obtained with the QMC method presented in the main text. The $L=4$ data match perfectly exact diagonalization results (solid lines).

\end{document}